\documentclass[12pt,oneside]{article}

\newcommand{\be}{\begin{equation}}
\newcommand{\bea}{\begin{eqnarray}}
\newcommand{\eea}{\end{eqnarray}}
\newcommand{\ba}{\begin{array}}
\newcommand{\ea}{\end{array}}
\newcommand{\ee}{\end{equation}}

\expandafter\ifx\csname mathbbm\endcsname\relax

\else

\fi \textheight 22cm \textwidth 15cm \topmargin 1mm \oddsidemargin 5mm \evensidemargin 5mm

\newcommand{\eqn}[1]{(\ref{#1})}

\def\Dslash{\,\,{\raise.15ex\hbox{/}\mkern-12mu D}}
\def\Dbarslash{\,\,{\raise.15ex\hbox{/}\mkern-12mu {\bar D}}}
\def\delslash{\,\,{\raise.15ex\hbox{/}\mkern-9mu \partial}}
\def\delbarslash{\,\,{\raise.15ex\hbox{/}\mkern-9mu {\bar\partial}}}
\def\pslash{\,\,{\raise.15ex\hbox{/}\mkern-9mu p}}
\def\calDslash{\,\,{\raise.15ex\hbox{/}\mkern-12mu {\cal D}}}

\def\theequation{\thesection.\arabic{equation}}

\input amssym.def
\input amssym.tex

\def\bbox{{\,\lower0.9pt\vbox{\hrule \hbox{\vrule height 0.2 cm
\hskip 0.2 cm \vrule height 0.2 cm}\hrule}\,}}

\def\Dslash{\,\,{\raise.15ex\hbox{/}\mkern-12mu D}}
\def\Dbarslash{\,\,{\raise.15ex\hbox{/}\mkern-12mu {\bar D}}}
\def\delslash{\,\,{\raise.15ex\hbox{/}\mkern-9mu \partial}}
\def\delbarslash{\,\,{\raise.15ex\hbox{/}\mkern-9mu {\bar\partial}}}
\def\pslash{\,\,{\raise.15ex\hbox{/}\mkern-9mu p}}
\def\calDslash{\,\,{\raise.15ex\hbox{/}\mkern-12mu {\cal D}}}

\begin{document}
\setcounter{page}{0}
\renewcommand{\theequation}{\thesection.\arabic{equation}}

\vspace{2cm}







\setcounter{equation}{0}

\begin{titlepage}

\begin{center}
{\today}
\hfill SLAC-PUB-11698\\
\hfill SU-ITP-06/05 \\

\vskip 1cm {\Large \bf Singularities and Closed String Tachyons \\} \vskip 1.3cm

{Eva Silverstein}\\

\vskip 1.3 cm

\vskip .2cm
{\sl ${}^2$SLAC and Department of Physics, Stanford University, \\
Stanford, CA 94309/94305, U.S.A. \\
{\tt evas@slac.stanford.edu} }

\end{center}

\vskip 0.2 cm
\begin{abstract}

A basic problem in gravitational physics is the resolution of spacetime singularities where general relativity
breaks down.  The simplest such singularities are conical singularities arising from orbifold identifications of
flat space, and the most challenging are spacelike singularities inside black holes (and in cosmology). Topology
changing processes also require evolution through classically singular spacetimes. I briefly review how a phase
of closed string tachyon condensate replaces, and helps to resolve, basic singularities of each of these types.
Finally I discuss some interesting features of singularities arising in the small volume limit of compact
negatively curved spaces and the emerging zoology of spacelike singularities.


\end{abstract}

\medskip
\medskip
\bigskip
\bigskip
\bigskip
\bigskip
\bigskip
\bigskip
\bigskip
\bigskip
\bigskip
\bigskip
\bigskip
\bigskip
\bigskip
\bigskip
\bigskip
\bigskip

\centerline{\it Based on comments at the 23rd Solvay conference in physics.}

\end{titlepage}

\section{Singularities and Winding Modes}

In the framework of string theory, several types of general relativistic singularities are replaced by a phase
of closed string tachyon condensate.  The simplest class of examples involves spacetimes containing
1-cycles with antiperiodic Fermion boundary conditions.   This class includes spacetimes which are globally
stable, such as backgrounds with late-time long-distance supersymmetry and/or AdS boundary conditions.

In the presence of such a circle, the spectrum of strings includes winding modes around the circle.  The Casimir
energy on the worldsheet of the string contributes a negative contribution to the mass squared, which is of the
form
\be M^2 = -{1\over l_s}^2 + {L^2\over l_s^4}  \ee
where $L$ is the circle radius and $1/l_s^2$ is the string tension scale.  For small $L$, the winding state
develops a negative mass squared and condenses, deforming the system away from the $L<l_s$ extrapolation of
general relativity. This statement is under control as long as $L$ is static or shrinking very slowly as it
crosses the string scale.  This sort of system was studied first in string theory in \cite{rohm}.

Examples include the following.  Generic orbifold singularities have twisted sector tachyons, {\it i.e.}
tachyons from strings wound around the angular direction of the cone.  The result of their condensation is that
the cone smooths out \cite{APS}, as seen in calculations of D-brane probes, worldsheet RG, and time dependent GR
in their regimes of applicability (see \cite{review}\ for reviews).  Topology changing transitions in which a
Riemann surface target space loses a handle or factorizes into separate surfaces are also mediated by winding
tachyon condensation \cite{TFA}. Tachyon condensation replaces certain spacelike singularities of a cosmological
type in which some number of circles shrinks homogeneously in the far past (or future) \cite{TE}.

Finally, tachyons condensing {\it quasilocally} over a spacelike surface appear in black hole problems and in a
new set of examples sharing some of their features \cite{newend}\cite{QLT}.  One interesting new example is an AdS/CFT dual
pair in which an infrared mass gap (confinement) arises at late times in a system which starts out in an
unconfined phase out on its approximate Coulomb branch.  As an example, consider the ${\cal N}=4$ SYM theory on
a (time dependent) Scherk-Schwarz circle, with scalar VEVs turned on putting it out on its Coulomb branch.  As
the circle shrinks to a finite size and the scalars roll back toward the origin, the infrared physics of the
gauge theory becomes dominated by a three dimensional confining theory.  The gravity-side description of this is
via a shell of D3-branes which enclose a finite region with a shrinking Scherk-Schwarz cylinder.  When the
cylinder's radius shrinks below the string scale, a winding tachyon turns on.  At the level of bulk spacetime
gravity, a candidate dual for the confining theory exists \cite{garymyers}; it is a type of ``bubble of nothing" in which the geometry smoothly caps off in the region corresponding to the infrared limit of the gauge theory.   This arises in the time dependent
problem via the tachyon condensate phase replacing the region of the geometry corresponding to the deep IR limit
of the field theory.

For all these reasons, it is important to understand the physics of the tachyon condensate phase.  The tachyon
condensation process renders the background time-dependent; the linearized solution to the tachyon equation of
motion yields an exponentially growing solution $T\propto \mu e^{\kappa X^0}$. As such there is no a priori preferred vacuum state. The
simplest state to control is a state $|out>$ obtained by a Euclidean continuation in the target space, and
describes a state in which nothing is excited in the far future when the tachyon dominates.   This is a
perturbative analogue of the Hartle-Hawking choice of state.  At the worldsheet level (whose self-consistency we
must check in each background to which we apply it), the tachyon condensation shifts the semiclassical action
appearing in the path integrand. String amplitudes are given by
\be <\prod\int V>\sim \int DX^0 D\vec X e^{-S_E} \prod \int V \label{pathint}\ee
where I work in conformal gauge and suppress the fermions and ghosts.  Here $X^\mu$ are the embedding
coordinates of the string in the target space and $\int V$ are the integrated vertex operators corresponding to
the bulk asymptotic string states appearing in the amplitude. The semiclassical action in the Euclidean theory
is
\be S_E=S_0 + \int d^2\sigma \mu^2 e^{2\kappa X^0} \hat T(\vec X) \label{action} \ee
with $S_0$ the action without tachyon condensation and $\hat T(\vec X)$ a winding (sine-Gordon) operator on the
worldsheet. These amplitudes compute the components of the state $|out>$ in a basis of multiple free string
states arising in the far past bulk spacetime when the tachyon is absent.  The tachyon term behaves like a
worldsheet potential energy term, suppressing contributions from otherwise singular regions of the path
integration.

Before moving to summarize the full calculation of basic amplitudes, let me note two heuristic indications that
the tachyon condensation effectively masses up degrees of freedom of the system.  First, the tachyon term in
\eqn{action}\ behaves like a spacetime dependent mass squared term in the analogue of this action arising in the
case of a first quantized worldline action for a relativistic particle \cite{StromTak}. Second, the dependence
of the tachyon term on the spatial variables $\vec X$ is via a relevant operator, dressed by worldsheet gravity
(which in conformal gauge is encapsulated in the fluctuations of the timelike embedding coordinate $X^0$).  The
worldsheet renormalization group evolution with scale is different from the time dependent evolution, since
fluctuations of $X^0$ contribute.  However in some cases, such as localized tachyon condensates and highly supercritical
systems, the two processes yield similar endpoints.  In any case, as a heuristic indicator of the effect of
tachyon condensation, the worldsheet RG suggests a massing up of degrees of freedom at the level of the
worldsheet theory as time evolves forward.

Fortunately we do not need to rely too heavily on these heuristics, as the methods of Liouville field theory
enable us to calculate basic physical quantities in the problem.  In the Euclidean state defined by the above
path integral, regulating the bulk contribution by cutting off $X^0$ in the far past at $ln\mu_*$, one finds a partition function $Z$ with real part
\be Re(Z)=-{ln(\mu/\mu_*)\over\kappa}\hat Z_{free} \label{partition}\ee
This is to be compared with the result from non-tachyonic flat space $Z_0=\delta(0)\hat Z_{free}$ \cite{TE},
where $\delta(0)$ is the infinite volume of time, and $\hat Z_{free}$ is the rest of the partition function. In
the tachyonic background \eqn{partition}, the first factor is replaced by a truncated temporal volume which ends
when the tachyon turns on.  A similar calculation of the two point function yields the Bogoliubov coefficients
corresponding to a pure state in the bulk with thermal occupation numbers of particles, with temperature
proportional to $\kappa$.  This technique was first suggested in \cite{StromTak}, where it was applied to bulk
tachyons for which $\kappa\sim 1/l_s$ and the resulting total energy density blows up. In the examples of
interest for singularities, the tachyon arises from a winding mode for which
$\kappa \ll 1/l_s$, and the method
\cite{StromTak}\ yields a self-consistently small energy density \cite{TE}.  In the case of an initial
singularity, this gives a perturbative string mechanism for the Hartle/Hawking idea of starting time from
nothing.  This timelike Liouville theory provides a perturbative example of ``emergent time", in the same sense
that spatial Liouville theory provides a worldsheet notion of ``emergent space".\footnote{This was also noted by
M. Douglas in a the discussion period in the session on emergent spacetime, in which G. Horowitz also noted
existing examples.   As explained by the speakers in that session, no complete {\it non-perturbative}
formulation involving emergent time exists, in contrast to the situation with spatial dimensions where matrix
models and AdS/CFT provide examples (but see \cite{craps}\ for an interesting
example of a null singularity with a proposed non-perturbative description in terms
of matrix theory).}

So far this analysis applied to a particular vacuum.  It is important to understand the status of other states
of the system.  In particular, the worldsheet path integral has a saddle point describing a single free
string sitting in the tachyon phase.  Do putative states such as this with extra excitations above the tachyon
condensate constitute consistent states?  This question is important for the problem of unitarity in black hole
physics and in more general backgrounds where a tachyon condenses quasilocally, excising regions of ordinary
spacetime. If nontrivial states persist in the tachyon phase in such systems, this would be tantamount to the
existence of hidden remnants destroying bulk spacetime unitarity.

In fact, we find significant indications that the state where a string sits by itself in the tachyon phase does
not survive as a consistent state in the interacting theory \cite{schomerus}\cite{QLT}.  The saddle point
solution has the property that the embedding coordinate $X^0$ goes to infinity in finite worldsheet time $\tau$.
This corresponds to a hole in the worldsheet, which is generically not BRST invariant by itself.  If mapped
unitarily to another hole in the worldsheet obtained from a correlated negative frequency particle impinging on
the singularity, worldsheet unitarity may be restored.  This prescription is a version of the Horowitz/Maldacena
proposal of a black hole final state \cite{finalstate}; the tachyon condensate seems to provide a microphysical
basis for this suggestion.  Note that this proposal is consistent with the idea of complimentarity:  unitary
evolution along the spacelike slice of the singularity is mapped to unitary evolution as seen by the asymptotic
observer.

A more dynamical effect which evacuates the tachyon region also
arises in this system.  A particle in danger of getting stuck in the
tachyon phase drags fields (for example the dilaton and graviton)
along with it.  The heuristic model of the tachyon condensate as an
effective mass for these modes \cite{StromTak}\ suggests that the
fields themselves are getting heavy.  The resulting total energy of
the configuration, computed in \cite{QLT}\ for a particle of initial
mass $m_0$ coupled with strength $\lambda$ to a field whose mass
also grows at late times like $M(x^0)$, is
\be E=m_0^2 \lambda^2 M(x^0) cos^2\biggl(\int^{x^0}M(t')dt'\biggr)
F(R) \ee
This is proportional to a function $F(R)$ which increases with
greater penetration distance $R$ of the particle into the tachyon
phase. Hence we expect a force on any configuration left in the
tachyon phase which sources fields (including higher components of
the string field). This does not mean every particle classically
gets forced out of the tachyonic sector:  for example in black hole
physics, the partners of Hawking particles which fall inside the
black hole provide negative frequency modes that correlate with the
matter forming the black hole.

The analysis of this dynamical effect in generic states
relies on the field-theoretic (worldsheet minisuperspace) model for tachyon dynamics.  It
is of interest to develop complete worldsheet techniques to analyze other putative vacua beyond the Euclidean
vacuum. In the case of the Euclidean vacuum, the worldsheet analyses \cite{StromTak}\cite{TE}\ reproduce
the behavior expected from the heuristic model, so we have tentatively taken it as a reasonable guide to the physics in more general
states as well.

The string-theoretic tachyon mode which drives the system away from the GR singularities is clearly accessible
perturbatively.  But it is important to understand whether the whole background has a self-consistent
perturbative string description. In the Euclidean vacuum, this seems to be the case:  the worldsheet amplitudes
are shut off in the tachyon phase in a way similar to that obtained in spatial Liouville theory. In other states, it is not a priori clear how far the perturbative treatment
extends.  One indication for continued perturbativity is that according to the simple field theory model, every
state gets heavy in the tachyon phase, including fluctuations of the dilaton, which may therefore be stuck at
its bulk weak coupling value.  It could be useful to employ AdS/CFT methods \cite{adscft}\ to help decide this point.

\section{Discussion and Zoology}

Many timelike singularities are resolved in a way that involves new
{\it light} degrees of freedom appearing at the singularity. In the
examples reviewed in section 1, ordinary spacetime ends where the
tachyon background becomes important.  The tachyon at first
constitutes a new light mode in the system, but its condensation
replaces the would-be short-distance singularity with a phase where
degrees of freedom ultimately become {\it heavy}. However, there are
strong indications that there is a whole zoo of possible behaviors
at cosmological spacelike singularities, including examples in which
the GR singularity is replaced by a phase with {\it more} light
degrees of freedom \cite{mutation}\ (see \cite{craps}\ for an
interesting null singularity where a similar behavior obtains).

In particular, consider a spacetime with compact negative curvature spatial slices, for example a Riemann
surface. The corresponding nonlinear sigma model is strongly coupled in the UV, and requires a completion
containing more degrees of freedom.  In  supercritical string theory, the dilaton beta function has a term
proportional to $D-D_{crit}$.  The corresponding contribution in a Riemann surface compactification is
$(2h-2)/V\sim 1/R^2$ where $V$ is the volume of the surface in string units, $h$ the genus and $R$ the curvature
radius in string units. This suggests that there are effectively $(2h-2)/V$ extra (supercritical) degrees of
freedom in the Riemann surface case. Interestingly, this count of extra degrees of freedom arises from the
states supported by the fundamental group of the Riemann surface.\footnote{I thank O. Aharony, A. Maloney, J.
McGreevy, and others for discussions on these points.}  For simplicity one can work at constant curvature and
obtain the Riemann surface as an orbifold of Euclidean $AdS_2$, and apply the Selberg trace formula to obtain
the asymptotic number density of periodic geodesics (as reviewed for example in \cite{BV}).  This yields a
density of states from a sum over the ground states in the winding sectors proportional to $e^{ml_s\sqrt{2h/V}}$
where $m$ is of order the mass of the string state. Another check arises by modular invariance which relates the
high energy behavior of the partition function to the lowest lying state: the system contains a light volume
mode, which is normalizable on the compact surface and propagates in loops, whose mass scales the right way to
account for the modular transform of this density of states.

At large radius, the Riemann surface component of the target space is clearly two dimensional to a good
approximation, and the 2d oscillator modes are entropically favored at high energy. It is interesting to
contemplate possibility of cases where the winding states persist to the limit $V\to 1$, in which case the
density of states from this sector becomes that of a $2h$ dimensional theory and the system crosses over to a
very supercritical theory in which these states become part of the oscillator spectrum. In particular, states
formed from the string wrapping generating cycles of the fundamental group in arbitrary orders (up to a small
number of relations) constitute a $2h$-dimensional lattice random walk.  At large volume, the lattice spacing is
much greater than the string scale and the system is far from its continuum limit, so these states are a small
effect.  But at small Riemann surface volume it is an interesting possibility that these states cross over to
the high energy spectrum of oscillator modes in $2h$ dimensions \cite{mutation}.

Of course as emphasized in \cite{mutation}, there are many possible behaviors at early times, including ones
where the above states do not persist to small radius \cite{TFA}\ and ones where they do persist but are part of
a still larger system. One simple way to complete the sigma model is to extend it to a linear sigma model
(containing more degrees of freedom) which flows to the Riemann surface model in the IR.  Coupling this system
to worldsheet gravity yields in general a complicated time dependent evolution, whose late time behavior is well
described by the nonlinear sigma model on an expanding Riemann surface.  If one couples this system to a large
supercritical spectator sector, then during the epoch when the coupling is weak (obtainable for any
semi-infinite time span by tuning the bare string coupling) the time dependent evolution approaches the RG flow
of the linear sigma model, which yields a controlled regime in which it is clear that at earlier times the
system had more degrees of freedom.

Clearly a priori this can happen in many ways.  In addition to the landscape of metastable vacua of string
theory I believe the conservative expectation is that there will be a zoo of possible cosmological histories
with similar late time behavior; indeed inflationary cosmology already has this feature.  While it may be
tempting to reject this situation out of hand in hopes of a universal prediction for all cosmological
singularities, that would be much more speculative.  Given the evidence for a plethora of solutions, there are
various indications that gravity may simplify in large dimensions (see e.g. \cite{largeD}) and it would be more
constructive to try to obtain from this an organizing principle or measure applying to the multitude of
cosmological singularities of this type.\footnote{as mentioned for example in J. Polchinski's talk in the
cosmology session.}

In any case, the singularities discussed in section 1, which are replaced by a phase of tachyon condensate, are
simpler, appear more constrained \cite{finalstate}\cite{QLT}, and apply more directly to black hole physics. It
would be interesting to understand if there is any principle relating black hole singularities and cosmological
singularities, and to try to apply these techniques to broad classes of GR singularities \cite{garfinkle}.

The relative simplicity with which the tachyon condensate removes the GR singularity provides a positive role
for tachyons; the instability which often motivates discarding systems with tachyons appears here in such a way
as to save the system from a worse (singular) fate. In fact even in low energy supersymmetric phenomenology it
is also not necessary to discard the whole space of models with bare negative scalar mass squareds
\cite{Tpheno}; it will be interesting to see if string compactifications can realize this predictive swath of
parameter space in a natural way \cite{JE}.

\medskip

\noindent{\bf Acknowledgements}

I would like to thank the organizers for a very interesting conference and for the invitation to present these
results and ideas in the session on singularities.  On these topics I have many people to thank, including my
collaborators A. Adams, O. Aharony, G. Horowitz, X. Liu, J. McGreevy, J. Polchinski, and A. Saltman. This
research is supported in part by the DOE under contract DE-AC03-76SF00515 and by the NSF under contract 9870115.

\end{document}